\documentclass[conference]{IEEEtran}
\IEEEoverridecommandlockouts

\usepackage{graphicx,color}
\usepackage{amsmath, amsthm, amsfonts, amssymb, amsbsy,nccmath,bbm}
\usepackage{algorithm}
\usepackage{algpseudocode}
\usepackage{enumerate}
\usepackage{lipsum}
\newtheorem{lemma}{Lemma}

\usepackage[sort,compress]{cite}
\usepackage{epsfig}
\usepackage{epstopdf}
\usepackage{mathtools}
\usepackage{dsfont}
\usepackage{epstopdf}

\usepackage{mathabx}

\usepackage[inline]{enumitem}   
\makeatletter
\newcommand{\inlineitem}[1][]{%
\ifnum\enit@type=\tw@
    {\descriptionlabel{#1}}
  \hspace{\labelsep}%
\else
  \ifnum\enit@type=\z@
       \refstepcounter{\@listctr}\fi
    \quad\@itemlabel\hspace{\labelsep}%
\fi}
\makeatother
\newtheorem{theorem}{Theorem}

\newcommand{\beq}{\begin{equation}}
\newcommand{\eeq}{\end{equation}}

\def\adots{\mathinner{\mskip0mu\raise0pt\vbox{\kern7pt\hbox{.}}\mskip3mu
          \raise4pt\hbox{.}\mskip3mu\raise8pt\hbox{.}\mskip0mu}}

\newcommand{\xh}{\widehat{x}}

\newcommand{\bmh}{\bfh}

\usepackage{bm}

\newcommand{\bmtheta}{{\bm \theta}}

\newcommand{\bme}{{\bm e}}
\newcommand{\bmx}{{\bm x}}
\newcommand{\bmy}{{\bm y}}
\newcommand{\bmv}{{\bm v}}

\newcommand{\bmH}{{\bm H}}

\renewcommand{\bmh}{{\bm h}}
\newcommand{\bmxh}{\widehat{\bmx}}

\newcommand{\bmA}{{\bm A}}

\newcommand{\bmC}{{\bf C}}

\newcommand{\bmD}{{\bf {D}}}

\newcommand{\bmzero}{{\bm 0}}
\newcommand{\bmp}{{\bm p}}
\newcommand{\bma}{{\bm a}}

\makeatletter
\newcommand\fs@spaceruled{\def\@fs@cfont{\bfseries}\let\@fs@capt\floatc@ruled
  \def\@fs@pre{\vspace{0.5\baselineskip}\hrule height.8pt depth0pt \kern2pt}%
  \def\@fs@post{\kern1pt\hrule\relax}%
  \def\@fs@mid{\kern2pt\hrule\kern2pt}%
  \let\@fs@iftopcapt\iftrue}
\makeatother

\newcommand{\bit}{\begin{itemize}}
\newcommand{\eit}{\end{itemize}}

\renewcommand{\bmh}{{\mathbf h}}
\renewcommand{\bmh}{{\mathbf h}}

\renewcommand{\bmA}{{\mathbf A}}

\newcommand{\bmB}{{\mathbf B}}

\newcommand{\bmr}{{\mathbf r}}

\usepackage{lipsum}

\newcommand{\bmXi}{{\boldsymbol \Xi}}
\newcommand{\bmxi}{{\boldsymbol \xi}}

\newcommand{\bmtau}{{\boldsymbol\tau}}
\renewcommand{\bmzero}{{\boldsymbol 0}}

\DeclareMathOperator{\E}{\mathbb{E}}
\newcommand{\var}{\mbox{var}}
\renewcommand{\bmxh}{\widehat{\bmx}}

\newcommand{\bmTheta}{{\bm{\Theta}}}

\newcommand{\ph}{\widehat{p}}
\newcommand{\bmmu}{{\bm{\mu}}}
\newcommand{\bmnu}{{\bm{\nu}}}

\newcommand{\nb}{{\overline{n}}}

\newcommand{\xc}{{\widecheck{x}}}

\newcommand{\taud}{{\dot{\tau}}}

\newcommand{\bmmud}{{\dot{\bmmu}}}
\newcommand{\bmCd}{{\dot{\bmC}}}

\newcommand{\mud}{{\dot{\mu}}}

\newcommand{\bd}{{\dot{b}}}
\newcommand{\bmtaud}{{\dot{\bmtau}}}

\usepackage[acronym]{glossaries}
\makeglossaries

\newacronym{kld}{KLD}{Kullback–Leibler divergence}

\newacronym{snr}{SNR}{signal-to-noise ratio}

\theoremstyle{remark}
\newtheorem*{remark}{Remark}

\begin{document}
\linespread{0.82}

\title{Expectations in Expectation Propagation
\vspace{-3mm}
}
\author{%
  \IEEEauthorblockN{Zilu Zhao, Fangqing Xiao, Dirk Slock}
  \IEEEauthorblockA{
			\small
			Communication Systems Department, EURECOM, France \\	
			\{zilu.zhao, fangqing.xiao, dirk.slock\}@eurecom.fr\vspace{-3mm}
					}
	}

\maketitle

\begin{abstract}
Expectation Propagation (EP) is a widely used message-passing algorithm that decomposes a global inference problem into multiple local ones. It approximates marginal distributions (beliefs) using intermediate functions (messages). While beliefs must be proper probability distributions that integrate to one, messages may have infinite integral values. In Gaussian-projected EP, such messages take a Gaussian form and appear as if they have "negative" variances. Although allowed within the EP framework, these negative-variance messages can impede algorithmic progress.

In this paper, we investigate EP in linear models and analyze the relationship between the corresponding beliefs. Based on the analysis, we propose both non-persistent and persistent approaches that prevent the algorithm from being blocked by messages with infinite integral values.

Furthermore, by examining the relationship between the EP messages in linear models, we develop an additional approach that avoids the occurrence of messages with infinite integral values.

\vspace{-2mm}
\end{abstract}

\section{Introduction}
\label{Intro}

Signal recovery is a fundamental problem in signal processing, with a wide range of applications.
In the Bayesian framework, however, canonical inference methods such as MMSE and MAP suffer from exponential computational complexity as the problem dimension increases.

Graphical-model-based iterative methods have proven effective by exploiting structural properties of the underlying models \cite{wainwright2008graphical}.
Expectation Propagation (EP), introduced in \cite{minka2005divergence}, transforms a global inference problem into multiple local inference problems.
EP can be interpreted as an iterative procedure for finding the stationary point of the constrained Bethe Free Energy (BFE) \cite{heskes2005approximate, zhang2021unifying}.
Within this framework, EP approximates marginal distributions (beliefs) using functions known as messages, which correspond one-to-one with the Lagrange multipliers in the constrained BFE formulation.
In EP and constrained BFE minimization, beliefs are required to be proper distributions—that is, they must be normalizable and integrate to one—whereas messages are allowed to have infinite integral values.

Although messages are not required to have finite integrals, messages with infinite integral values may produce beliefs that also fail to integrate to a finite value and may even prevent the algorithm from progressing.

\subsection{Main Contribution}
The revisited Vector Approximate Message Passing (reVAMP) algorithm \cite{EURECOM+7412} is an instance of EP applied to linear models, whose joint probability density function factorizes into a likelihood term and symbol-wise priors. 
In this paper, we analyze the messages and beliefs in reVAMP and show that the belief at the likelihood factor node is guaranteed to have a finite integral value whenever the beliefs at the prior nodes have finite integral values.

Based on this observation, we propose both non-persistent and persistent approaches that prevent EP/reVAMP from being blocked by messages with negative variances, while still allowing such messages to exist when they do not hinder the algorithm.

Furthermore, we show that the messages to the prior factors always have finite integral values whenever the messages to the likelihood factor have finite integral values. In addition, we investigate the Kullback–Leibler divergence (KLD) minimization step in EP and propose Analytic Continuation reVAMP (ACreVAMP), which prevents the occurrence of messages with negative variances.

\subsection{Notations}
We use the notation $[\bmH]_{i, j}$ to denote the entry in the $i$-th row and $j$-th column $\bmH$. Following MATLAB convention, we denote $[\bmH]_{:, i}$ as the $i$-th column of $\bmH$, and for simplicity we write $\bmh_{i}$ as shorthand for $[\bmH]_{:, i}$.

Similarly, $[\bmtheta]_{i}$ denotes the $i$-th entry of the vector $\bmtheta$, and we write $\theta_i$ as 
shorthand for $[\bmtheta]_{i}$.

We define the operator $\text{Diag}(\bmtheta)$, which maps a vector $\bmtheta$ to a diagonal matrix $\bmTheta=\text{Diag}(\bmtheta)$ such that $[\bmTheta]_{i, i}=\theta_{i}$.
Conversely, the operator $\text{diag}(\bmTheta)$ extracts the diagonal entries of an arbitrary matrix $\bmTheta$ into a vector $\bmtheta=\text{diag}(\bmTheta)$, where $[\bmtheta]_i=[\bmTheta]_{i, i}$.

Finally, we refer to a non-negative function as a proper distribution if it integrates to one. Throughout the paper, we normalize all messages and beliefs to one whenever their integrals are finite.

\section{Expectation Propagation}
Consider a factored joint PDF with non-negative factors (denoted as $\forall \alpha:\, f_{\alpha}$):
\beq
    p(\bmtheta)\propto \prod_{\alpha}f_{\alpha}(\bmtheta_{\alpha}),
    \label{eq:lkasdjf}
\eeq
where for each $\alpha$, $f_{\alpha}(\bmtheta_\alpha)$ has finite integral value, and the vector $\bmtheta_{\alpha}$ denotes a sub-vector of $\bmtheta$. The goal of EP is to approximate the marginal statistics (e.g. mean and variance) of the distribution in \eqref{eq:lkasdjf}. 

A corresponding factor graph for \eqref{eq:lkasdjf} is constructed by introducing a factor node for every factor $f_{\alpha}$ and a variable node for every variable $\theta_i$ in the vector $\bmtheta$. The graph is then completed by connecting each factor node to every variable node that appears as a parameter in that factor (the connecting line is called an edge).

Each edge carries two types of messages: a variable-to-factor message and a factor-to-variable message.
Furthermore, each factor and variable node forms a belief that is uniquely determined by the incoming messages. These beliefs serve as approximations to the marginals of $p(\bmtheta)$.

\begin{remark}
In EP, messages are not required to be proper distributions and may integrate to infinity, whereas beliefs must be proper distributions and must always normalize to one.
For simplicity, throughout this paper we normalize even the messages to one whenever they have finite integral value.
\end{remark}

EP iteratively updates all messages using the following rules:
\begin{itemize}
    \item Belief at the factor node, $\forall \alpha:$
    \beq
        \begin{split}
             b_{f_{\alpha}}(\bmtheta_{\alpha})\propto f_{\alpha}(\bmtheta_{\alpha})\prod_{i\in N(\alpha)} \Delta^{f_{\alpha}\leftarrow \theta}_{i}(\theta_i).
        \end{split}
        \label{eq:lasdkjf}
    \eeq
    \item Message from factor to variable, $ \forall \alpha, i\in N(\alpha):$
    \beq
        \Delta^{f_{\alpha}\to \theta}_{i}(\theta_i)\propto \frac{\text{proj}[\int  b_{f_{\alpha}}(\bmtheta_{\alpha}) \prod_{i'\in N(\alpha)/
        \{i\}}d\theta_{i'}]}{\Delta^{f_{\alpha}\leftarrow \theta}_{i}(\theta_i)}.
        \label{eq:asdfjbdfn}
    \eeq
    \item Belief at the variable node, $\forall i:$
    \beq
        b_{\theta_{i}}(\theta_i)\propto \prod_{\alpha\in N(i)} \Delta^{f_{\alpha}\to \theta}_{i}(\theta_i).
        \label{eq:ljfdsb}
    \eeq
    \item Message from variable to factor $\forall i, \alpha\in N(i):$
    \beq
        \Delta^{f_{\alpha}\leftarrow \theta}_{i}(\theta_i)\propto \frac{b_{\theta_{i}}(\theta_i)}{\Delta^{f_{\alpha}\to \theta}_{i}(\theta_i)}.
        \label{eq:ajdgbkjfn}
    \eeq
\end{itemize}
In \eqref{eq:lasdkjf}–\eqref{eq:ajdgbkjfn}, $N(i)$ denotes the set of factor indices whose corresponding factors are directly connected to variable $\theta_i$ (i.e., in the neighborhood of $\theta_i$). Similarly, $N(\alpha)$ denote the set of variable indices connected to factor $f_{\alpha}$.

In \eqref{eq:asdfjbdfn}, the operator $\text{proj}[\cdot]$ projects an arbitrary distribution onto a desired family by minimizing the Kullback–Leibler divergence (KLD). 
In this paper, we consider only the Gaussian family $\mathcal{G}$. Thus,
\beq
    \text{proj}[p(\theta)]\triangleq \arg\min_{q(\theta)\in \mathcal{G}} KLD[p(\theta)\|q(\theta)].
    \label{eq:lnfjbk}
\eeq
If $\mathcal{G}$ is Gaussian, minimizing \eqref{eq:lnfjbk} is equivalent to matching the mean and variance of $p(\theta)$ and $q(\theta)$. We use the term "extrinsic of factor $f_{\alpha}$" to denote the product of incoming messages to $f_{\alpha}$, such as the message product in \eqref{eq:lasdkjf}.

\subsection{Problem Formulation}
Observe the message in \eqref{eq:asdfjbdfn}. 
The division in this update may yield a function with an infinite integral value even when both the numerator and denominator are proper distributions.
Consequently, the resulting message may appear to be a Gaussian with a "negative" variance.
Although such messages are allowed within EP, they can hinder the algorithm because the beliefs in \eqref{eq:lasdkjf} and \eqref{eq:ljfdsb} are required to remain Gaussian.

In the remainder of this paper, we examine this negative-variance issue in linear models and propose several approaches to prevent the resulting blocking behavior.

\section{Useful Relations}
In this section, we list several useful relations that will be used throughout the paper.
\subsection{Matrix Inversion Lemma}
\beq
    \bmD(\bmA+\bmB\bmC\bmD)^{-1}=\bmC^{-1}(\bmD\bmA^{-1}\bmB+\bmC^{-1})^{-1}\bmD\bmA^{-1
    }.\nonumber
\eeq
\subsection{Matrix Determinant Lemma}
\beq
\det(\bmA+\bmB\bmC\bmD)=\det(\bmA)\det(\bmC)\det(\bmC^{-1}+\bmD\bmA^{-1}\bmB).
\label{eq:asilomar2529}
\eeq
\subsection{Multivariate Gaussian Reproduction Lemma}
\beq
    \begin{split}
        &\mathcal{N}(\bmx|\bmmu_1, \bmC_1)\mathcal{N}(\bmA\bmx|\bmmu_2, \bmC_2)\\
        &=\mathcal{N}(\bmmu_2|\bmA\bmmu_1, \bmA\bmC_1\bmA^T+\bmC_2)\mathcal{N}(\bmx|\bmmu_3, \bmC_3),
    \end{split}
    \label{eq:asilomar2502}
\eeq
where
\beq
\begin{split}
    \bmC_3&=(\bmC_1^{-1}+\bmA^T\bmC_2^{-1}\bmA)^{-1};\\
    \bmmu_3&=\bmC_3(\bmC_1^{-1}\bmmu_1+\bmA^T\bmC_2^{-1}\bmmu_2).
\end{split}
\label{eq:asilomar2503}
\eeq

\section{System Model}
We consider the linear measurement model:
\beq
    \bmy=\bmA\bmx+\bmv,
    \label{eq:sysmodel}
\eeq
where $\bmA\in \mathbb{R}^{M\times N}$, the noise $\bmv\sim\mathcal{N}(\bmzero, \bmC_\bmv)$, and the signal vector $\bmx$ has independent entries, i.e., $\forall n\in [1, N]:\, x_n\sim p(x_n)$. 
The joint PDF associated with the system model \eqref{eq:sysmodel} factorizes as
\beq
    p(\bmx, \bmy)=f_{\bmy}(\bmx)\prod_{n}f_{x_n}(x_n),
    \label{eq:fskajbkjsdanv}
\eeq
with factors
\beq
    \begin{split}
        &f_{\bmy}(\bmx)\triangleq p(\bmy|\bmx)=\mathcal{N}(\bmy|\bmA\bmx, \bmC_{\bmv})\\
        &f_{x_n}(x_n)\triangleq p(x_n).
    \end{split}
    \label{eq:jdgjkbcbcbcc}
\eeq

\section{Brief Introduction to reVAMP}
\label{sect:asilomar25VI}
The explicit forms of the messages in reVAMP are summarized in Table~\ref{table:1},  
where we denote
\beq
    \mathcal{UN}(\bmtheta|\bmmu, \bmC)\triangleq\exp{-\frac{1}{2}(\bmtheta-\bmmu)^T\bmC^{-1}(\bmtheta-\bmmu)}.
\eeq
\begin{table}[ht!]
\centering
\begin{tabular}{||c | c | c||} 
 \hline
\!\!Message and Beliefs\!\! & Notation& Explicit Form (proportional) \\ [0.5ex] 
 \hline\hline
 $x_{n}\to f_{\bmy}$ & $\Delta^{f_\bmy \leftarrow x}_{n}(x_n, t)$ & $\mathcal{UN}(x_n| \mu_{p_n}(t), \tau_{p_n}(t))$ \\ 
 $f_{x_n} \to x_n$ & $\Delta^{f_{x} \to x}_{n}(x_n, t)$ & $\mathcal{UN}(x_n| \mu_{p_n}(t), \tau_{p_n}(t))$ \\
 $f_{\bmy} \to x_n$ & $\Delta^{f_{\bmy} \rightarrow x}_{n}(x_n, t)$ & $\mathcal{UN}(x_n| \mu_{r_n}(t), \tau_{r_n}(t))$\\
 $x_n \to f_{x_n}$ & $\Delta^{f_{x}\leftarrow x}_{n}(x_n, t)$ & $\mathcal{UN}(x_n| \mu_{r_n}(t), \tau_{r_n}(t))$\\
 Belief at $f_\bmy$ & $b_{f_{\bmy}}(\bmx, t)$ & $f_{\bmy}(\bmx)\prod_{n}\!\!\Delta^{f_\bmy \leftarrow x}_{n}(x_n, t)$ \\
 Belief at $f_{x_n}$ & $b_{f_{x_n}}(x_n, t+1)$ & $f_{x_n}(x_n)\Delta^{f_{x} \leftarrow x}_{n}(x_n, t)$ \\[1ex]
 \hline
\end{tabular}
\vspace{1mm}
\caption{Explicit Forms of Messages and Beliefs}
\label{table:1}
\end{table}

\begin{figure}[htb]
\centerline{\includegraphics[width=0.48\textwidth]{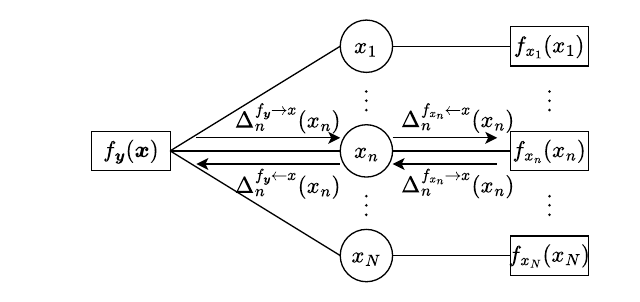}}
\vspace{-3mm}
\caption{Sparse Signal Recovery}
\label{fig:factorgraph}
\vspace{-3mm}
\end{figure}

The factor graph corresponding to the joint PDF \eqref{eq:fskajbkjsdanv} is illustrated in Fig.~\ref{fig:factorgraph}.

In EP and reVAMP, messages are not required to be normalized to one and may even have infinite integral values. In contrast, beliefs must be proper distributions whose integrals equal one. For simplicity, throughout this paper we normalize the messages and beliefs in Table~\ref{table:1} to one whenever their integrals are finite.

To make the discussion concise, we introduce the following vector and matrix notations:
\[
    \bmmu_{\bmp}(t) \!\!=\!\! 
    \begin{bmatrix}
        \mu_{p_1}(t) \!\!&\!\! \dots \!\!&\!\! \mu_{p_N}(t)
    \end{bmatrix}^T, \, 
    \bmtau_{\bmp}(t) \!\!= \!\!
    \begin{bmatrix}
        \tau_{p_1}(t) \!\!&\!\! \dots \!\!&\!\! \tau_{p_N}(t)
    \end{bmatrix}^T,
\]
and define
\[
    \bmC_{\bmp}(t) = \mathrm{Diag}(\bmtau_{\bmp}(t)).
\]
Similarly, we stack $\mu_{r_n}(t)$ and $\tau_{r_n}(t)$ into vectors $\bmmu_{\bmr}(t)$ and $\bmtau_{\bmr}(t)$, respectively.

Let $\bma_n$ denote the $n$-th column of $\bmA$, and let the subscript $\nb$ indicate that the $n$-th column or entry is removed. For example,
    \beq
    \begin{split}
    &\bmA_{\nb}\triangleq\begin{bmatrix}
        \bma_1 &\dots &\bma_{n-1} &\bma_{n+1} &\dots \bma_{N}
    \end{bmatrix}.
    \end{split}
    \nonumber
    \eeq
Furthermore, we define
\beq
    \begin{split}
    &\bmmu_{\bmp_{\nb}}(t)\!\triangleq\! \begin{bmatrix}
        \mu_{p_1}(t)\!\! &\!\!\dots \!\!&\!\!\mu_{p_{n-1}}(t) \!\!&\!\! \mu_{p_{n+1}}(t) \!\!&\!\!\dots \!\!&\!\!\mu_{p_{N}}(t)
    \end{bmatrix}^T\\
    &\bmtau_{\bmp_{\nb}}(t)\!\!\triangleq\!\begin{bmatrix}
        \tau_{p_1}(t)\!\! &\!\!\dots \!\!&\!\!\tau_{p_{n-1}}(t) \!\!&\!\! \tau_{p_{n+1}}(t) \!\!&\!\!\dots \!\!&\!\!\tau_{p_{N}}(t)
    \end{bmatrix}^T\\
    &\bmC_{\bmp_{\nb}}(t+1)\triangleq\text{Diag}[\bmtau_{\bmp_{\nb}}(t+1)].
    \end{split}
    \label{eq:labelforbar}
\eeq

In addition to the above, we use the shorthand $\forall n$ to denote $\forall n \in \{1,\dots, N\}$.

Throughout this paper, unless explicitly stated otherwise (for example, in the proof of Lemma \ref{th:asilomarlm1}), we distinguish between the index $n$ and the time-dependent index $n(t)$. The index $n$ refers to an arbitrary entry, whereas $n(t)$ can be viewed as a selection function denoting the entry selected at time $t$. The two indices are independent of each other; $n$ has no relation to $n(t)$, and vice versa. To avoid ambiguity, we explicitly use the quantifiers $\forall$ and $\exists$ whenever needed.

Next, we briefly describe the ideal reVAMP procedure, assuming that no negative message variances occur.

\subsection{Ideal reVAMP: Message from $f_{\bmy}$ to $\bmx$\label{sec:lkdsnfdbfnf}}

The belief at the likelihood factor can be written as
\beq
\begin{split}
    b_{f_\bmy}(\bmx, t)&\propto f_{\bmy}(\bmx)\prod_{n}\Delta^{f_\bmy \leftarrow x}_{n}(x_n, t)\\
    &\propto \mathcal{N}(\bmx|\bmmu_{\bmxh|\bmy}(t), \bmC_{\bmxh|\bmy}(t)),
\end{split}
\label{eq:EP50}
\eeq
where
\beq
\begin{split}
&\bmC_{\bmxh|\bmy}(t)=\left(\bmA^T\bmC_{\bmv}^{-1}\bmA+\bmC_{\bmp}(t)^{-1}\right)^{-1}\\
&\bmmu_{\bmxh|\bmy}(t)=\bmC_{\bmxh|\bmy}(t)\left(\bmA^T\bmC_{\bmv}^{-1}\bmy+\bmC_{\bmp}(t)^{-1}\bmmu_{\bmp}(t)\right).
\end{split}
\label{eq:asilomar2538}
\eeq
Since \eqref{eq:EP50} is already Gaussian, its marginals are also Gaussian, and the KLD projection has no effect. Denote the $n$-th diagonal entry of $\bmC_{\bmxh|\bmy}(t)$ by $\tau_{\xh_n|\bmy}(t)=\bme_{n}^T\bmC_{\bmxh|\bmy}(t)\bme_{n}$, and the $n$-th entry of $\bmmu_{\bmxh|\bmy}(t)$ by $\mu_{\xh_n|\bmy}(t)=\bme_n^T\bmmu_{\bmxh|\bmy}(t)$. Then the message from $f_{\bmy}$ to $x_n$ is
\beq
\begin{split}
    \forall n:\, \Delta^{f_{\bmy} \rightarrow x}_{n}(x_n, t)&\propto \frac{\text{proj}(\int b_{f_{\bmy}}(\bmx, t) \prod_{n'\neq n}dx_{n'})}{\Delta^{f_\bmy \leftarrow x}_{n}(x_n, t)}\\
    &\propto \mathcal{UN}(x_n| \mu_{r_n}(t), \tau_{r_n}(t)),
\end{split}
\label{eq:asilomarmessagerori}
\eeq
where $\forall n$:
\beq
    \begin{split}
        &\tau_{r_{n}}(t)=\left(\frac{1}{\tau_{\xh_n|\bmy}(t)}-\frac{1}{\tau_{p_n}(t)}\right)^{-1}\\
        &\mu_{r_n}(t)=\tau_{r_{n}}(t)\left(\frac{\mu_{\xh_n|\bmy}(t)}{\tau_{\xh_n|\bmy}(t)}-\frac{\mu_{p_n}(t)}{\tau_{p_n}(t)}\right).
    \end{split}
    \label{eq:asilomar2541ta}
\eeq
As we can see, the updates of $\tau_{r_n}(t)$ and $\mu_{r_{n}}(t)$ for all $n$ depend only on the values of $\bmmu_{\bmp}(t)$ and $\bmtau_{\bmp}(t)$. Using the matrix inversion lemma, one can verify that \eqref{eq:asilomar2541ta} is equivalent to the following expressions (i.e., both \eqref{eq:asilomar2541ta} and \eqref{bprime} yield the same results):
\beq
    \begin{split}
        &\forall n: \tau_{r_{n}}(t)=F_{\tau_{r_n}}\left[\bmmu_{\bmp_{\nb}}(t), \bmtau_{\bmp_{\nb}}(t)\right]\\
        &\forall n: \mu_{r_n}(t)=F_{\mu_{r_n}}\left[\bmmu_{\bmp_{\nb}}(t), \bmtau_{\bmp_{\nb}}(t)\right],
    \end{split}
    \tag{\ref{eq:asilomar2541ta}$'$}\label{bprime}
\eeq
where $\forall n:$
\beq
\begin{split}
    F_{\tau_{r_n}}[\bmmu_{\bmp_{\nb}}(t), \bmtau_{\bmp_{\nb}}(t)]\triangleq\left[\bma_n^T(\bmC_\bmv+\bmA_{\nb}\bmC_{\bmp_{\nb}}(t)\bmA_{\nb})^{-1}\bma_n\right]^{-1}\\
    F_{\mu_{r_n}}\left[\bmmu_{\bmp_{\nb}}(t), \bmtau_{\bmp_{\nb}}(t)\right]\triangleq\left[\bma_n^T(\bmC_\bmv+\bmA_{\nb}\bmC_{\bmp_{\nb}}(t)\bmA_{\nb})^{-1}\bma_n\right]^{-1}\\
    \cdot\bma_{n}^T(\bmC_\bmv+\bmA_{\nb}\bmC_{\bmp_{\nb}}(t)\bmA_{\nb})^{-1}(\bmy-\bmA_{\nb}\bmp_{\nb}(t)).
\end{split}
\label{eq:asilomar2516ss}
\eeq

\subsection{Ideal reVAMP: Message from $f_{x_{n(t)}}$ to $x_{n(t)}$}
\label{sec:asilomar25subsec2}
Since we employ sequential updates, suppose that at computation time $t$, we are updating the pair $(\mu_{p_{n(t)}}(t+1), \tau_{p_{n(t)}}(t+1))$, where $n(t)$ is a selection function that chooses an index in $\{1,\dots, N\}$ based on time $t$. 

According to the EP rules, the belief at the prior factor $f_{x_{n(t)}}$ is given by
\beq
    b_{f_{x_{n(t)}}}(x_{n(t)}, t+1)\propto f_{x_{n(t)}}(x_{n(t)})\Delta^{f_{x}\leftarrow x}_{n(t)}(x_{n(t)}, t).
    \label{eq:asilomar2541}
\eeq
To obtain the mean and variance of $b_{f_{x_{n(t)}}}$, a non-analytic (numerical) evaluation may be required, since $f_{x_{n(t)}}(x_{n(t)})=p(x_{n(t)})$ defined in \eqref{eq:jdgjkbcbcbcc} can be an arbitrary prior. Nevertheless, this evaluation is only one-dimensional. We define the mean and variance of belief in \eqref{eq:asilomar2541} as:
\beq
    \begin{split}
        &\mu_{\xh_{n(t)}|\bmmu_\bmr}(t+1)\triangleq\E_{b_{f_{x_{n(t)}}}(t+1)}[x_{n(t)}]\\
        &\tau_{\xh_{n(t)}|\bmmu_\bmr}(t+1)\triangleq\var_{b_{f_{x_{n(t)}}}(t+1)}[x_{n(t)}].
    \end{split}
    \label{eq:asilomar2523}
\eeq
Again following the EP rules, the outgoing message from the prior factor is
\beq
\begin{split}
    \Delta^{f_{x} \to x}_{n(t)}(x_{n(t)}, t+1)\propto \frac{\text{proj}[b_{f_{x_{n(t)}}}(x_{n(t)}, t+1)]}{\Delta^{f_{x}\leftarrow x}_{n(t)}(x_{n(t)}, t)}\\
    \propto \mathcal{UN}(x_{n(t)}|\mu_{p_{n(t)}}(t+1), \tau_{p_{n(t)}}(t+1)),
    \label{eq:asilomar2544}
\end{split}
\eeq
where
\beq
    \begin{split}
        &\tau_{p_{n(t)}}(t\!+\!1)\!=\!\left(\frac{1}{\tau_{\xh_{n(t)}|\bmmu_\bmr}(t+1)}-\frac{1}{\tau_{r_{n(t)}}(t)}\right)^{-1};\\
        &\mu_{p_{n(t)}}(t\!+\!1)\!=\!\tau_{p_{n(t)}}(t+1)\left(\frac{\mu_{\xh_{n(t)}|\bmmu_\bmr}(t+1)}{\tau_{\xh_{n(t)}|\bmmu_\bmr}(t+1)}-\frac{\mu_{r_{n(t)}}(t)}{\tau_{r_{n(t)}}(t)}\right).
    \end{split}
    \label{eq:asilomar45}
\eeq
For all other indices $\forall n'\neq n(t)$, the parameters remain unchanged:
\beq
    \forall n' \!\neq\! n(t):\, \mu_{p_{n'}}(t+1)\!=\!\mu_{p_{n'}}(t), \tau_{p_{n'}}(t+1)\!=\!\tau_{p_{n'}}(t)
    \label{eq:asilomar2546ss}
\eeq
This closes the reVAMP update loop. By iteratively updating the pairs $(\bmmu_{\bmp}, \bmtau_{\bmp})$ and $(\bmmu_{\bmr}, \bmtau_{\bmr})$ until convergence, we obtain an approximation of the posterior mean: $\E[\bmx|\bmy]$ as $\E[\bmx|\bmy]\simeq \bmmu_{\bmxh|\bmy}(\infty)\simeq \bmmu_{\bmxh|\bmmu_\bmr}(\infty)$.

\subsection{Low-Rank Update}
It is worth noting that the update of $\bmC_{\bmxh|\bmy}$ requires only $O(N^2)$ multiplications if only one diagonal entry of $\bmC_{\bmp}$ is updated. Since the pair $(\mu_{p_{n(t)}}, \tau_{p_{n(t)}})$ is updated at time $t$, we define
\beq
    \frac{1}{\Delta\tau_{p_{n(t)}}(t+1)}\triangleq\frac{1}{\tau_{p_{n(t)}}(t+1)}-\frac{1}{\tau_{p_{n(t)}}(t)}.
\eeq
Then the rank-one update for $\bmC_{\bmxh|\bmy}$ is 
\beq
\begin{split}
\bmC_{\bmxh|\bmy}(t+1)&=\bmC_{\bmxh|\bmy}(t)-[\bmC_{\bmxh|\bmy}(t)]_{:, n(t)}\cdot\\
&\cdot (\Delta\tau_{p_{n(t)}}(t+1)+\tau_{\xh_{n(t)}|\bmy}(t))^{-1}[\bmC_{\bmxh|\bmy}(t)]_{:, n(t)}^T.
\end{split}
\label{eq:asilomar2548}
\eeq

Note that in Section~\ref{sec:lkdsnfdbfnf}, it may appear that we compute the parameters $(\mu_{r_n}(t), \tau_{r_n}(t))$ for all $n$ in a parallel fashion. However, this is effectively equivalent to computing only $(\mu_{r_{n(t)}}(t), \tau_{r_{n(t)}}(t))$, because the parameters corresponding to the other indices are not used in Section~\ref{sec:asilomar25subsec2}. Nevertheless, expressing the update of $(\mu_{r_n}(t), \tau_{r_n}(t))$ in parallel form simplifies the discussions in Section~\ref{sec:asilomarVII} and \ref{sect:jdfuab}. Moreover, the low-rank update of the belief covariance still requires $O(N^2)$ operations, whereas computing all $\forall n: (\mu_{r_n}(t), \tau_{r_n}(t))$ incurs only $O(N)$ complexity.

Although EP permits messages that are not proper distributions—such as Gaussian-like messages with negative variances—these messages may induce beliefs that are no longer proper distributions. This is not allowed in EP and may prevent the algorithm from progressing. For example, if $\tau_{r_{n(t)}}(t)$ is negative, the belief in \eqref{eq:asilomar2541} may fail to be a proper distribution, and the expectation operations in \eqref{eq:asilomar2523} may become ill-defined.

\section{Keeping Negative Variance}
\label{sec:asilomarVII}
In EP, the messages in Table~\ref{table:1} are not required to be proper distributions; in particular, their variances may take negative values. In contrast, the beliefs $b_{f_{\bmy}}$ and $\forall n: b_{f_{x_n}}$ must be proper distributions. 
From the previous discussion, evaluating $b_{f_{\bmy}}$ in \eqref{eq:EP50} in \eqref{eq:EP50} involves only Gaussian PDF manipulations, whereas obtaining the statistics of $\forall n: b_{f_{x_n}}$ in \eqref{eq:asilomar2541} requires integrating with respect to an arbitrary prior. Therefore, we propose to allow negative values of $\tau_{p_n}$ and apply special treatment only when a given $\tau_{p_n}$ would lead to a negative value of some $\tau_{r_n'}$.

\begin{lemma}
\label{th:asilomarlm1}
If, in reVAMP, sequential updates are used and $\tau_{p_{n(t)}}(t+1)$ is updated via \eqref{eq:asilomar45}, then
    \beq
        \frac{\det[\bmC_{\bmxh|\bmy}(t)]}{\det[\bmC_{\bmxh|\bmy}(t+1)]}=\frac{\tau_{\xh_{n(t)|\bmy}}(t)}{\tau_{\xh_{n(t)|\bmmu_\bmr}}(t+1)}.
    \eeq    
\end{lemma}
\begin{proof}
    Recall from Section~\ref{sect:asilomar25VI} that $n(t)$ denotes the index of the entry in $\bmmu_{\bmp}$ and $\bmtau_{\bmp}$ that is updated at computation time $t$. For brevity, in this proof (from \eqref{eq:asilomar47} to \eqref{eq:asilomar2555sss}), we use the shorthand $n = n(t)$.

     Define
    \beq
    \begin{split}
        \bmC_{\bmxh_{\nb}|\bmy}(t+1)^{-1}\triangleq\bmC_{\bmxh|\bmy}(t+1)^{-1}-\bme_{n}\tau_{p_n}(t+1)^{-1}\bme_{n}^T\\
        =\bmC_{\bmxh|\bmy}(t)^{-1}-\bme_{n}\tau_{p_n}(t)^{-1}\bme_{n}^T.
        \label{eq:asilomar47}
    \end{split}
    \eeq
    Using \eqref{eq:asilomar2529} and \eqref{eq:asilomar47}, we obtain 
    \beq
        \begin{split}
            \det(\bmC_{\bmxh|\bmy}(t+1)^{-1})=\det(\bmC_{\bmxh_{\nb}|\bmy}(t+1)^{-1})\\
            \cdot\frac{\tau_{p_n}(t+1)+\bme_{n}^T\bmC_{\bmxh_{\nb}|\bmy}(t+1)\bme_{n}}{\tau_{p_n}(t+1)}.
        \end{split}
        \label{eq:asilomar2549}
    \eeq
    Since, under sequential updates at computation time $t$, only $\tau_{\bmp_{n}}(t+1)$ is updated while all other entries remain unchanged, we have
    \beq
    \begin{split}
        &\text{Sequential Update }\Rightarrow\,\forall n'\neq n(t): \tau_{p_{n'}}(t+1)=\tau_{p_{n'}}(t+1)\\
        &\Rightarrow \bmtau_{\bmp_{\nb}}(t+1)=\bmtau_{\bmp_{\nb}}(t)\\
        &\Rightarrow \bmC_{\bmp_{\nb}}(t+1)=\bmC_{\bmp_{\nb}}(t).
    \end{split}
    \label{eq:discussion111}
    \eeq
    From \eqref{eq:discussion111}, and using the matrix inversion lemma, we obtain
    \beq
    \begin{split}
        \bme_{n}^T\bmC_{\bmxh_{\nb}|\bmy}(t+1)\bme_{n}=\tau_{r_{n}}(t).
        \label{eq:asilomar2550}
    \end{split}
    \eeq

    Substituting the update expression for $\tau_{p_n}(t+1)$ in \eqref{eq:asilomar45} together with \eqref{eq:asilomar2550} into \eqref{eq:asilomar2549} yields
     \beq
        \begin{split}
            \det(\bmC_{\bmxh|\bmy}(t+1)^{-1})=\det(\bmC_{\bmxh_{\nb}|\bmy}(t+1)^{-1})\frac{\tau_{r_{n}}(t)}{\tau_{\xh_n|\bmmu_\bmr}(t+1)}.
        \end{split}
        \label{eq:asilomar2549ss}
    \eeq

    Next, from the second line of \eqref{eq:asilomar47} and the matrix determinant lemma \eqref{eq:asilomar2529}, we have
    \beq
    \begin{split}
        \det(\bmC_{\bmxh_{\nb}|\bmy}(t+1)^{-1})=\det(\bmC_{\bmxh|\bmy}(t)^{-1})\frac{\tau_{p_n}(t)-\tau_{\xh_n|\bmy}(t)}{\tau_{p_n}(t)}\\
        =\det(\bmC_{\bmxh|\bmy}(t)^{-1}) \tau_{\xh_n|\bmy}(t) \left(\frac{1}{\tau_{\xh_n|\bmy}(t)}-\frac{1}{\tau_{p_n}(t)}\right)
        \label{eq:asilomar2553}
    \end{split}
    \eeq
    Substituting the right-hand side of the first line of \eqref{eq:asilomar2541ta} into \eqref{eq:asilomar2553}, and then inserting the resulting expression into \eqref{eq:asilomar2549ss}, we obtain
     \beq
        \begin{split}
            &\det(\bmC_{\bmxh|\bmy}(t+1)^{-1})=\det(\bmC_{\bmxh|\bmy}(t)^{-1})\frac{\tau_{\xh_n|\bmy}(t)}{\tau_{\xh_n|\bmmu_\bmr}(t+1)}.\\
            &\Leftrightarrow \frac{\det[\bmC_{\bmxh|\bmy}(t)]}{\det[\bmC_{\bmxh|\bmy}(t+1)]}=\frac{\tau_{\xh_{n(t)|\bmy}}(t)}{\tau_{\xh_{n(t)|\bmmu_\bmr}}(t+1)}.
        \end{split}
        \label{eq:asilomar2555sss}
    \eeq
\end{proof}

\begin{theorem}
    In reVAMP as described in Section \ref{sect:asilomar25VI}, if $\forall t:, b_{f_{x_n(t)}}(x_{n(t)}, t+1)$ in \eqref{eq:asilomar2541} are proper distributions (normalize to one) with respect to $x_{n(t)}$, and if $\bmtau_{\bmp}(1)$ is initialized with positive entries, then the belief covariance matrices $\forall t:\, \bmC_{\bmxh|\bmy}(t)$ in \eqref{eq:asilomar2538} remain positive definite.
    \label{th:asilomar251}
\end{theorem}
\begin{proof}
    We use mathematical induction together with Sylvester's criterion.
    
    Since we initialize $\forall n:\, \tau_{p_n}(1)$ with positive numbers, the matrix $\bmC_{\bmxh|\bmy}(1)$ at the beginning of the first iteration is positive definite.

    We can always reorder the entries of $\bmx$ and the corresponding columns of $\bmA$ such that the updated $\tau_{p_n}(t+1)$ appears in the last diagonal entry of $\bmC_{\bmxh|\bmy}(t+1)^{-1}$. 
    Thus, without loss of generality, assume that $N=n(t)$, and that $\bmC_{\bmxh|\bmy}(t)$ is positive definite (induction hypothesis). We will show that $\bmC_{\bmxh|\bmy}(t+1)$ is also positive definite.

    By Sylvester's criterion, a symmetric matrix is positive definite if and only if all its leading principal minors (the determinants of its upper-left $k \times k$ submatrices for all $k$) are positive. The only difference between $\bmC_{\bmxh|\bmy}(t)^{-1}$ and $\bmC_{\bmxh|\bmy}(t+1)^{-1}$ is the last diagonal entry. 
   Hence, the first through $(N-1)$-th leading principal minors of $\bmC_{\bmxh|\bmy}(t+1)^{-1}$ and $\bmC_{\bmxh|\bmy}(t)^{-1}$ coincide and are therefore positive. It remains to show that the $N$-th order leading principal minor, i.e., $\det(\bmC_{\bmxh|\bmy}(t+1)^{-1})$, is also positive.

    From {Lemma \ref{th:asilomarlm1}}, with the assumption that $N=n(t)$, we have
     \beq
        \begin{split}
            \det(\bmC_{\bmxh|\bmy}(t+1)^{-1})=\det(\bmC_{\bmxh|\bmy}(t)^{-1})\frac{\tau_{\xh_N|\bmy}(t)}{\tau_{\xh_N|\bmmu_\bmr}(t+1)}.
        \end{split}
        \label{eq:asilomar2555}
    \eeq

    By the induction hypothesis, $\det(\bmC_{\bmxh|\bmy}(t)^{-1})>0$ and $\tau_{\xh_N|\bmy}(t)>0$. Moreover, from \eqref{eq:asilomar2523} and the assumption that $b_{f_{x_{n(t)}}}(x_{n(t)}, t+1)$ is a proper distribution, we have $\tau_{\xh_N|\bmmu_\bmr}(t+1)>0$. Hence, $\det(\bmC_{\bmxh|\bmy}(t+1)^{-1})>0$. Since all leading principal minors of $\bmC_{\bmxh|\bmy}(t+1)^{-1}$ are positive, the matrix is positive definite.
\end{proof}

\begin{theorem}
\label{cor:asilomar251}
    If the initial values $\forall n:\, \tau_{p_{n}}(1)$ are all positive, then the covariance matrix $\forall t:\, \bmC_{\bmxh|\bmy}(t)$ remains positive definite when using any mixture of the following two update strategies for $\forall t:\, \tau_{p_{n(t)}}(t+1)$:
    \begin{itemize}
        \item $\tau_{p_{n(t)}}(t+1)$ is updated via \eqref{eq:asilomar45}
        \item $\tau_{p_{n(t)}}(t+1) := \tau_{p_{n(t)}}(t)$
    \end{itemize}
\end{theorem} 
\begin{proof}
    We again use mathematical induction. As in Theorem~\ref{th:asilomar251}, $\bmC_{\bmxh|\bmy}(1)$ is positive definite by assumption. We need to show that $\bmC_{\bmx|\bmy}(t+1)$ is positive definite given that $\bmC_{\bmx|\bmy}(t)$ is positive definite. 

    If $\tau_{p_{n(t)}}(t+1)$ is updated via \eqref{eq:asilomar45}, we can follow the same proof as in {Theorem~\ref{th:asilomar251}} using {Lemma~\ref{th:asilomarlm1}} to conclude that $\bmC_{\bmxh|\bmy}(t+1)$ is positive definite.

    If instead $\tau_{p_{n(t)}}(t+1)$ is set to $\tau_{p_{n(t)}}(t+1) := \tau_{p_{n(t)}}(t)$, then $\bmC_{\bmp}(t+1)=\bmC_{\bmp}(t)$, since for entries with indices $\forall n'\neq n(t)$, the update rule \eqref{eq:asilomar2546ss} implies that $\tau_{p_{n'}}(t+1) = \tau_{p_{n'}}(t)$.
    Consequently, $\bmC_{\bmxh|\bmy}(t+1)=\bmC_{\bmxh|\bmy}(t)$ which is positive definite by the induction hypothesis.
\end{proof}

\begin{remark}
    Theorem~\ref{th:asilomar251} implies that, at any stationary point, the belief $b_{f_\bmy}(\bmx, \infty)$ automatically becomes a proper distribution if $\forall n: b_{f_{x_n}}(x_n, \infty)$ are proper distributions. During the iterations, however, Theorem~\ref{cor:asilomar251} relaxes the assumption in Theorem~\ref{th:asilomar251} that  $b_{f_{x_n(t)}}(x_{n(t)}, t+1)$ must integrate to one for all $t$, and it provides an alternative mechanism for handling message updates when $b_{f_{x_n}}$ is not a proper distribution. From a practical standpoint, Theorems~\ref{th:asilomar251} and~\ref{cor:asilomar251} offer two complementary strategies for dealing with negative message variances.
\end{remark}

In the following, we propose two simple augmentations of reVAMP to handle messages with negative variances. Based on Theorem~\ref{th:asilomar251}, we propose an update scheme with an explicit checking procedure (non-persistent approach). Based on Theorem~\ref{cor:asilomar251}, we also propose a persistent scheme without such checking.

\subsection{Non-Persistent Approach}
According to Theorem~\ref{th:asilomar251}, we must ensure that, at every time $t$, the selection function $n(t)$ chooses an index for which the corresponding $b_{f_{x_{n(t)}}}(x_{n(t)}, t+1)$, obtained from \eqref{eq:asilomar2541} is a proper distribution. Furthermore, since EP can be interpreted as an iterative method for finding stationary points of the constrained BFE, we know that, at convergence, all beliefs must be proper distributions.

Motivated by these two considerations, we propose to check the values $\forall n, \tau_{r_n}(t+1)$ when using \eqref{eq:asilomar45} to compute $\tau_{p_{n(t)}}(t+1)$ from $\tau_{p_{n(t)}}(t)$, before actually applying the update to $\tau_{p_{n(t)}}(t)$.

\subsubsection{\label{sec:subsubstart}} During each iteration $t$ of this non-persistent EP scheme, we first follow \eqref{eq:EP50}-\eqref{eq:asilomar2523}.

\subsubsection{} Next, we compute the unchecked updates in the same manner as in \eqref{eq:asilomar45}, but denote them with a dot to distinguish them from the checked values:
\beq
    \begin{split}
        &\taud_{p_{n(t)}}(t\!+\!1)=\left(\frac{1}{\tau_{\xh_{n(t)}|\bmmu_\bmr}(t+1)}-\frac{1}{\tau_{r_{n(t)}}(t)}\right)^{-1};\\
        &\mud_{p_{n(t)}}(t\!+\!1)\!=\!\taud_{p_{n(t)}}(t\!+\!1)\left(\frac{\mu_{\xh_{n(t)}|\bmmu_\bmr}(t\!+\!1)}{\tau_{\xh_{n(t)}|\bmmu_\bmr}(t\!+\!1)}\!-\!\frac{\mu_{r_{n(t)}}(t)}{\tau_{r_{n(t)}}(t)}\right).
    \end{split}
    \label{eq:asilomar2531}
\eeq
Meanwhile $\forall n'\neq n(t)$, we set
\beq
\mud_{p_{n'}}(t+1)=\mu_{p_{n'}}(t),\, \taud_{p_{n'}}(t+1)=\tau_{p_{n'}}(t)
\eeq
As a result, 
\beq
\begin{split}
    &\bmtaud_{\bmp}(t+1)\!=\!\bmtau_{\bmp}(t)+\bme_{n(t)}[\taud_{p_{n(t)}}(t+1)-\tau_{p_{n(t)}}(t)]\\
    &\bmCd_{\bmp}(t+1)\!=\!\bmC_{\bmp}(t)+\bme_{n(t)}[\taud_{p_{n(t)}}(t+1)-\tau_{p_{n(t)}}(t)]\bme_{n(t)}^T\\
    &\bmmud_{\bmp}(t+1)\!=\!\bmmu_{\bmp}(t)+\bme_{n(t)}[\mud_{p_{n(t)}}(t+1)-\mu_{p_{n(t)}}(t)].
\end{split}
\nonumber
\eeq

\subsubsection{}
We then check whether this unchecked update would lead to beliefs that are not proper distributions.
To this end, we first compute the extrinsic messages of the prior factors. Based on \eqref{eq:asilomar2541ta} and \eqref{bprime}, we obtain $\forall n$:
\beq
    \begin{split}
        &\taud_{r_{n}}(t+1)=F_{\tau_{r_n}}[\bmmud_{\bmp_{\nb}}(t+1), \bmtaud_{\bmp_{\nb}}(t+1)]\\
        &\mud_{r_n}(t+1)=F_{\mu_{r_n}}[\bmmu_{\bmp_{\nb}}(t+1), \bmtaud_{\bmp_{\nb}}(t+1)],
    \end{split}
    \label{eq:asilomar2541tasds}
\eeq
where, for all $n$ and all $n'\neq n$, the vectors $\bmmud_{\bmp_{\nb}}(t+1)$ and $\bmtaud_{\bmp_{\nb}}(t+1)$ stack the elements $\mu_{p_{n'}}(t+1)$ and $\tau_{p_{n'}}(t+1)$, respectively, following the same convention as in \eqref{eq:labelforbar}.

Then, according to \eqref{eq:asilomar2541}, the beliefs resulting from the unchecked update $\mud_{p_{n(t)}}(t+1)$ and $\taud_{p_{n(t)}}(t+1)$  are
\beq
\begin{split}
    \forall n\!:\!\bd_{f_{x_n}}(x_n, t\!+\!2)\propto f_{x_n}(x_n)\mathcal{UN}(x_n| \mud_{r_n}(t\!+\!1), \taud_{r_{n}}(t\!+\!1)).
    \label{eq:asilomar2541sdfsd}
\end{split}
\eeq

\subsubsection{} 
Finally, the actual (checked) updates are defined as
\beq
\begin{split}
    &\tau_{p_{n(t)}}(t+1)\\
    &=\begin{cases}
        \taud_{p_{n(t)}}(t+1)\, & \forall n: \int \bd_{f_{x_{n}}}(x_{n}, t+2) \text{d}x_{n}<\infty\\
        \tau_{p_{n(t)}}(t)\, &\text{otherwise}
    \end{cases}\\
    &\mu_{p_{n(t)}}(t+1)\\
    &=\begin{cases}
        \mud_{p_{n(t)}}(t+1)\, & \forall n: \int \bd_{f_{x_{n}}}(x_{n}, t+2) \text{d}x_{n}<\infty\\
        \mu_{p_{n(t)}}(t)\, &\text{otherwise}, 
    \end{cases}
\end{split}
\label{eq:klasjdfl}
\eeq
where we use the notation $\int \text{d}f(\theta)<\infty$ to denote that the function $f(\theta)$ has finite integral value. 
If evaluating the integrals $\forall n: \int \bd_{f_{x_{n}}}(x_{n}, t+2) \text{d}x_{n}$ in \eqref{eq:klasjdfl} is too costly, we may instead check the signs of $\forall n: \taud_{r_{n}}(t+1)$ in \eqref{eq:asilomar2541tasds}. This leads to a relaxed non-persistent approach:
\begin{itemize}
\item 
If $\forall n: \taud_{r_{n}}(t+1)> 0$, then $\tau_{p_{n(t)}}(t+1)$ and $\mu_{p_{n(t)}}(t+1)$ are updated to $\taud_{p_{n(t)}}(t+1)$ and $\mud_{p_{n(t)}}(t+1)$ . 
\item 
If $\exists n: \tau_{r_{n}}(t+1)\leq 0$, then $\tau_{p_{n(t)}}(t+1)$ and $\mu_{p_{n(t)}}(t+1)$ remain unchanged as $\tau_{p_{n(t)}}(t)$ and $\mu_{p_{n(t)}}(t)$.
\end{itemize}
We refer to the scheme based on \eqref{eq:klasjdfl} as the strict non-persistent approach, and to the sign-based rule above as the {relaxed} non-persistent approach.

\subsubsection{}Afterward, we proceed to iteration $t+1$ and repeat from Section \ref{sec:subsubstart}.

\subsection{Persistent Approach}

In this approach, we do not perform any look-ahead. At computation time $t$, we proceed from \eqref{eq:EP50} to \eqref{eq:asilomar2516ss} following the standard EP/reVAMP steps to obtain $\forall n: \mu_{r_{n}}(t)$ and $\tau_{r_{n}}(t)$. 

After that, before updating $\tau_{p_{n(t)}}(t+1)$ and $\mu_{p_{n(t)}}(t+1)$, we check whether $b_{f_{x_{n(t)}}}(x_{n(t)}(t), t+1)$, obtained from \eqref{eq:asilomar2541}, is a proper distribution (integrates to finite value).

\begin{itemize}
    \item If the belief $b_{f_{x_{n(t)}}}(x_{n(t)}(t), t+1)$ is a proper distribution, then $\tau_{p_{n(t)}}(t+1)$ and $\mu_{p_{n(t)}}(t+1)$ are updated by following \eqref{eq:asilomar2541}-\eqref{eq:asilomar2546ss}.
    \item Otherwise if the belief $b_{f_{x_{n(t)}}}(x_{n(t)}(t), t+1)$ is not a proper distribution, then $\tau_{p_{n(t)}}(t+1)$ and $\mu_{p_{n(t)}}(t+1)$ remain unchanged, i.e., 
    \beq
    \forall n\!\in\!\{1,\dots, N\}\!:\! \mu_{p_{n}}(t\!+\!1)\!\!=\!\!\mu_{p_{n}}(t),\, \tau_{p_{n}}(t\!+\!1)\!\!=\!\!\tau_{p_{n}}(t).
    \label{eq:asilomar2546sgdfgs}
    \eeq
\end{itemize}
If checking whether $b_{f_{x_{n(t)}}}(x_{n(t)}(t), t+1)$ is a proper distribution via \eqref{eq:asilomar2541} is computationally demanding, we can relax the condition and instead check the sign of $\tau_{r_{n(t)}}$. If $\tau_{r_{n(t)}}$ is positive, we proceed with the standard update in \eqref{eq:asilomar2541}-\eqref{eq:asilomar2546ss}; otherwise, we apply the no-update rule in \eqref{eq:asilomar2546sgdfgs}. 

\section{Avoiding Negative Message Variances}
\label{sect:jdfuab}
By matrix inversion lemma (analog to \eqref{eq:asilomar2550}), one can show that the messages  $\mathcal{UN}(\bmx| \bmmu_{\bmr}(t), \text{Diag}[\bmtau_\bmr(t)])$ are proper distributions if $\forall n, t:\, \tau_{p_{n}}(t)$  are positive. Therefore, in this section we focus on the message from $f_{x_n}$ to $x_n$ and on enforcing $\forall n, t:\, \tau_{p_n}(t)>0$. W.l.o.g., in the following of this section, we assume the Ansatz that $\forall n, t: \, \tau_{r_n}(t)>0$.

To gain further insight into \eqref{eq:asilomar2541}, we abstract a simplified one-dimensional measurement model whose posterior PDF shares the same factorization structure as that in \eqref{eq:asilomar2541}:
\beq
    \mu_r=x+v_r, \label{eq:asilomar2555ss}
\eeq
where $v_r\sim\mathcal{N}(0, \tau_{r})$, and $x\sim p(x)$ is an arbitrary distribution. 
We denote the posterior PDF, mean and variance of \eqref{eq:asilomar2555ss} as $b_{f_x}(x)=p(x|\mu_r)$,  $\mu_{\xh|\mu_r}$ and $\tau_{\xh|\mu_r}$.
According to the EP rule, negative message variances occur when the posterior variance $\tau_{\xh|\mu_r}$ is larger than $\tau_{r}$.

\subsection{A Simple Example of Negative Message Variances}
Suppose $p(x)$ is BPSK. Negative message variances arise when we observe an incorrect measurement with high precision. A simple example is $\mu_r=0$ and $\tau_r$  arbitrarily small, because in the BPSK case with $\mu_r=0$, the posterior variance $\tau_{\xh|\mu_r}$ is independent of $\tau_r$.

\subsection{Reconsidering the KLD Projection}
\label{sec:asilomarsubB}
Based on EP rule and \eqref{eq:asilomar2544}, the message parameterized by the pair $(\mu_p, \tau_p)$ is updated by
\beq
\begin{split}
    &(\mu_{\xc|\mu_r}, \tau_{\xc|\mu_r})=\arg\!\!\!\!\!\!\!\!\min_{\mu_{\xc|\mu_r}, \tau_{\xc|\mu_r}}\!\!\!\! KLD[b_{f_x}(x)\|\mathcal{N}(x|\mu_{\xc|\mu_r}, \tau_{\xc|\mu_r})];\\
    &\mathcal{UN}(x|\mu_p, \tau_p)\propto \frac{\mathcal{N}(x|\mu_{\xc|\mu_r}, \tau_{\xc|\mu_r})}{\mathcal{N}(x|\mu_r, \tau_r)}.
\end{split}
\label{eq:asilomar2556}
\eeq
Consider the second line of \eqref{eq:asilomar2556} and impose the constraint $\tau_p> 0$. Then, we have
\beq
    \mathcal{N}(x|\mu_{\xc|\mu_r}(p, \tau_p), \tau_{\xc|\mu_r}(p, \tau_p))\propto \mathcal{N}(x|\mu_r, \tau_r) \mathcal{N}(x|\mu_p, \tau_p),
    \nonumber
\eeq
where we explicitly emphasize the dependence of $(\mu_{\xc|\mu_r}, \tau_{\xc|\mu_r})$ on the unknown
parameters $(\mu_p, \tau_p)$:
\beq
\begin{split}
    &\tau_{\xc|\mu_r}(\mu_p, \tau_p)=\frac{\tau_p\tau_r}{\tau_p+\tau_r};\\
    &\mu_{\xc|\mu_r}(\mu_p, \tau_p)=\frac{\tau_p \mu_r + \tau_r \mu_p}{\tau_p+\tau_r}.
\end{split}
\label{eq:asilomar2558}
\eeq
Substituting \eqref{eq:asilomar2558} into the first line of \eqref{eq:asilomar2556} and imposing the constraint $\tau_{p}> 0$ yields the following constrained optimization problem:
\begin{equation}
\begin{aligned}
\min_{\mu_p, \tau_p}  \quad & KLD[b_{f_x}(x)\|\mathcal{N}(x|\mu_{\xc|\mu_r}(\mu_p, \tau_p), \tau_{\xc|\mu_r}(\mu_p, \tau_p))],\\
\textrm{s.t.} \quad & \tau_p > 0.    \\
\end{aligned}
\label{eq:asilomar2559}
\end{equation}
Note that \eqref{eq:asilomar2559} may yield a different solution from \eqref{eq:asilomar2556}
due to the additional constraint on $\tau_{p}$.

Expanding the KLD, \eqref{eq:asilomar2559} can be rewritten as
\begin{equation}
\begin{aligned}
\min_{\mu_p, \tau_p}  \quad & L(\mu_p, \tau_p)\\
\textrm{s.t.} \quad & \tau_p> 0,    \\
\end{aligned}
\label{eq:asilomar2560}
\end{equation}
where
\beq
L(\mu_p, \tau_p)\!\triangleq\!\frac{\ln{\tau_{\xc|\mu_r}(\mu_p, \tau_p)}}{2}+\frac{\tau_{\xh|\mu_r}\!\!+\![\mu_{\xh|\mu_r}\!\!-\!\mu_{\xc|\mu_r}(\mu_p, \tau_p)]^2}{2\tau_{\xc|\mu_r}(\mu_p, \tau_p)}.
\eeq
Taking the partial derivative of the objective function with respect to $\mu_p$ and setting it to zero
yields
\beq
    \frac{\partial L(\mu_p, \tau_p)}{\partial \mu_p}=0\Rightarrow \mu_{\xc|\mu_r}(\mu_p, \tau_p)=\mu_{\xh|\mu_r}
    \label{eq:asilomar2562}
\eeq
Next, compute the partial derivative with respect to $\tau_p$ and substitute \eqref{eq:asilomar2562} into the result:
\beq
    \left.\frac{\partial L(\mu_p, \tau_p)}{\partial \tau_p}\right|_{\mu_{\xc|\mu_r}(\mu_p, \tau_p)=\mu_{\xh|\mu_r}}=\frac{\xi_p^2(\xi_{\ph}-\xi_p)}{\frac{2}{\tau_{\xh|\mu_r}}(\xi_r+\xi_p)},
    \label{eq:asilomar2563}
\eeq
where we define $\xi_p\triangleq\frac{1}{\tau_p}$, $\xi_r\triangleq\frac{1}{\tau_r}$, and $\xi_{\ph}\triangleq\frac{1}{\tau_{\xh|\mu_r}}-\frac{1}{\tau_r}$. From these definitions, we have $\xi_{\ph}> \xi_r$ and $\xi_r> 0$. Thus, we only need to consider two cases: $\xi_{\ph}> 0$ and $\xi_{\ph}\leq 0$.
\subsubsection{If $\xi_{\ph}> 0$}
In this case, the optimum is attained at $\xi_{p}=\xi_{\ph}$. Therefore, the update rule in \eqref{eq:asilomar45} remains unchanged.
\subsubsection{If $\xi_{\ph}\leq 0$}
In this case, \eqref{eq:asilomar2563} is monotonically increasing in $\xi_p>0$, and therefore the minimum is achieved as $\xi_{p}\to 0$, which corresponds to $\tau_p\to\infty$. Instead of computing the corresponding mean, we inspect the Gaussian natural parameters and define $\nu_{p}=\frac{\mu_p}{\tau_p}=\xi_p\mu_p$. Substituting $\tau_p\to \infty$ into \eqref{eq:asilomar2562}, we obtain $\nu_p\to \frac{\mu_{\xh|\mu_r}-\mu_r}{\tau_r}$.
\subsection{Analytic Continuation of ReVAMP}
We now revisit the discussion in Section~\ref{sect:asilomar25VI}. To extend the message domain of reVAMP, $\forall n$ we propose to use Gaussian natural parameters
\beq
\begin{split}
    \xi_{p_n}(t)=\tau_{p_n}(t)^{-1};\;
    \nu_{p_n}(t)=\frac{\mu_{p_n(t)}}{\tau_{p_n}(t)}
\end{split}
\eeq
instead of the $(\mu_{p_n}(t), \tau_{p_n}(t))$ representation of the messages.
Furthermore, we define 
\beq
\begin{split}
    &\bmxi_{\bmp}(t)=\text{diag}[\bmC_{\bmp}(t)^{-1}]=\begin{bmatrix}
        \xi_{p_1}(t) &\dots & \xi_{p_N}(t)
    \end{bmatrix}^T\\
    &\bmXi_{\bmp}(t)=\bmC_{\bmp}(t)^{-1}=\text{Diag}[\bmxi_{\bmp}(t)];\\
    &\bmnu_{\bmp}(t)=\bmC_{\bmp}(t)^{-1}\bmmu_{\bmp}(t).
\end{split}
\eeq
The extended reVAMP algorithm with analytic continuation (ACreVAMP) is summarized as follows.

\subsubsection{ACreVAMP Message from $f_{\bmy}$ to $\bmx$}
To incorporate infinite variances, we rewrite the belief mean and covariance in \eqref{eq:EP50} as
\beq
\begin{split}
&\bmC_{\bmxh|\bmy}(t)=\left(\bmA^T\bmC_{\bmv}^{-1}\bmA+\bmXi_{\bmp}(t)\right)^{-1}\\
&\bmmu_{\bmxh|\bmy}(t)=\bmC_{\bmxh|\bmy}(t)\left(\bmA^T\bmC_{\bmv}^{-1}\bmy+\bmnu_{\bmp}(t)\right).
\end{split}
\label{eq:asilomar2566}
\eeq
Similar to Section \ref{sect:asilomar25VI}, we denote, $\forall n,$ the $n$-th diagonal entry in $\bmC_{\bmx|\bmy}$ as $\tau_{\xh_n|\bmy}$ and $\forall n,$ the $n$-th entry of $\bmmu_{\bmxh|\bmy}$ as $\mu_{\xh_n|\bmy}$. Extending the EP rule to allow infinite $\tau_{p_n}(t)$, the message update in \eqref{eq:asilomarmessagerori}-\eqref{eq:asilomar2541ta} becomes, $\forall n,$
\beq
    \begin{split}
         \tau_{r_{n}}(t)=\left(\frac{1}{\tau_{\xh_n|\bmy}(t)}-\xi_{p_n}(t)\right)^{-1}\\
        \mu_{r_n}(t)=\tau_{r_{n}}(t)\left(\frac{\mu_{\xh_n|\bmy}(t)}{\tau_{\xh_n|\bmy}(t)}-\nu_{p_n}(t)\right).
    \end{split}
    \label{eq:asilomar2567a}
\eeq

\subsubsection{ACreVAMP Message from $f_{x_{n}}$ to $x_{n}$}
Unlike the methods in Section \ref{sec:asilomarVII}, ACreVAMP can be carried out in both parallel and sequential forms. In this paper, we focus on the sequential update to maintain compatibility with the previous discussions.

As discussed earlier, $\forall n, t$: $\tau_{r_n}(t)$  is positive if $\forall n, t : \tau_{p_n}(t)$ is positive. Therefore, the computation of the belief mean $\mu_{\xh_{n(t)}|\bmmu_\bmr}(t+1)$ and variance $\tau_{\xh_{n(t)}|\bmmu_\bmr}(t+1)$ remains the same as in \eqref{eq:asilomar2541}-\eqref{eq:asilomar2523}.

However, based on the discussion in Section \ref{sec:asilomarsubB}, the parameters $\nu_{p_{n(t)}}(t+1)$ and $\xi_{p_{n(t)}}(t+1)$ are updated as:
\beq
\begin{split}
    &\nu_{p_{n(t)}}\!(t+1)\!\!=\!\!\begin{cases}
    \nu_{\ph_{n(t)}}(t+1)\, & \xi_{\ph_{n(t)}}(t+1)> 0\\
    \frac{\mu_{\xh_{n(t)}|\bmmu_{\bmr}}(t+1)-\mu_{r_{n(t)}}(t)}{\tau_{r_{n(t)}}(t)} \, & \text{otherwise}
    \end{cases}\\
    &\xi_{p_{n(t)}}(t+1)=\begin{cases}
        \xi_{\ph_{n(t)}}(t+1) \quad & \xi_{\ph_{n(t)}}(t+1)> 0\\
        0 \quad & \text{otherwise,}
    \end{cases}
\end{split}
\nonumber
\eeq
where
\beq
    \begin{split}
        &\xi_{\ph_{n(t)}}(t+1)=\frac{1}{\tau_{\xh_{n(t)}|\bmmu_\bmr}(t+1)}-\frac{1}{\tau_{r_{n(t)}}(t)}\\
        &\nu_{\ph_{n(t)}}(t+1)=\frac{\mu_{\xh_{n(t)}|\bmmu_\bmr}(t+1)}{\tau_{\xh_{n(t)}|\bmmu_\bmr}(t+1)}-\frac{\mu_{r_{n(t)}}(t)}{\tau_{r_{n(t)}}(t)}
    \end{split}.
\eeq
\subsubsection{Low Rank Update}
If sequential update is used, we can still reduce the complexity by exploiting matrix inversion lemma. Define $\Delta\xi_{p_{n(t)}}(t+1)\triangleq\xi_{p_{n(t)}}(t+1)-\xi_{p_{n(t)}}(t)$. Analogous to \eqref{eq:asilomar2548}, the belief covariance matrix is updated as
\beq
\bmC_{\bmxh|\bmy}(t\!+\!1)\!\!=\!\!
\begin{cases}
    \!\bmC_{\bmxh|\bmy}(t)\!+\!\Delta\bmC_{\bmxh|\bmy}(t\!+\!1), \, & \Delta\xi_{p_{n(t)}}(t\!+\!1)\neq 0\\
    \!\bmC_{\bmxh|\bmy}(t),\, & \text{otherwise}
\end{cases}
\eeq
where, for the non-trivial cases $\Delta\xi_{p_{n(t)}}(t+1)\neq 0$
\beq
\begin{split}
    &\Delta\bmC_{\bmxh|\bmy}(t+1)=-[\bmC_{\bmxh|\bmy}(t)]_{:, n(t)}\cdot\\
    &\cdot(\Delta\xi_{p_{n(t)}}(t+1)^{-1}+\tau_{\xh_{n(t)}|\bmy}(t))^{-1}[\bmC_{\bmxh|\bmy}(t)]_{:, n(t)}^T
\end{split}
\eeq
This closes the loop of ACreVAMP.

\section{Numerical Simulations}
In our simulations, we consider two scenarios: sparse signal recovery and binary phase-shift keying (BPSK) signal recovery. In both cases, i.i.d. white Gaussian noise is assumed.
\subsection{Sparse Signal Recovery Setup}
To verify the proposed approach in sparse signal recovery, we consider a measurement matrix $\bmA \in \mathbb{R}^{M \times N}$ of size $M \times N = 8 \times 10$. Each entry of $\bmA$ is i.i.d. and drawn from a Gaussian distribution, i.e., for all $m, n$, $[\bmA]_{m, n}\sim \mathcal{N}(0, 1/N)$. The signal vector $\bmx$ contains independent but non-identically distributed Gaussian-mixture entries with exponentially decaying amplitudes:
\beq
\begin{split}
    \forall n: p(x_n)=0.5\mathcal{N}(x_n|-1\times3.2^{1-n}, 0.1\times 3.2^{2-2n})\\
    +0.5\mathcal{N}(x_n|1\times3.2^{1-n}, 0.1\times 3.2^{2-2n}).
\end{split}
\eeq

\subsection{BPSK Signal Recovery Setup}

In the BPSK signal recovery scenario, we consider systems of size $M \times N = 20 \times 10$.
Similarly, the entries of the measurement matrix $\bmA$ are i.i.d. Gaussian: $\forall m, n: [\bmA]_{m, n}\sim \mathcal{N}(0, 1/N)$. BPSK can be viewed as a Gaussian mixture whose components have (ideally) zero variance. In our simulations, we approximate this by the following i.i.d. signal prior:
\beq
\begin{split}
    \forall n: p(x_n)=0.5\mathcal{N}(x_n|-1, 0.01)+0.5\mathcal{N}(x_n|1, 0.01).
\end{split}
\nonumber
\eeq

\subsection{Simulation Results}
In both scenarios, we vary the noise power and evaluate the algorithms at SNR levels $\forall i=\{1,\dots, 11\}: SNR=5(i-1) \text{dB}$. For each SNR level, $500$ independent problem instances are generated. We denote by $\mathcal{I}_{l}$ the set of indices corresponding to all instances simulated at $SNR=l$.

We use the normalized mean squared error (NMSE) relative to the MMSE solution as the performance metric. At SNR level $SNR=l$, the NMSE is defined as
\beq
    NMSE_l=\frac{\sum_{i\in \mathcal{I}_{l}}(\xh_i-\bmxh_{\text{MMSE}, i})^T(\xh-\bmxh_{\text{MMSE}, i})}{\sum_{i\in \mathcal I_l}\bmxh_{\text{MMSE}, i}^T\bmxh_{\text{MMSE}, i}}.
\eeq
The exact MMSE result is computed by brute-force evaluation with complexity $\mathcal{O}(2^N N^3)$. 

We compare the proposed methods against LMMSE, the widely used clipping approach \cite{10378663}, and Noise-Unbiasing
\cite{9053763}. The clipping approach is conceptually similar to the discussion in
Section~\ref{sect:jdfuab}, in that it aims to avoid negative message variances. However, unlike the proposed ACreVAMP, if $\tau_{p_{n(t)}}(t+1)$ obtained from \eqref{eq:asilomar45} is non-positive, the clipping method resets the message variance to $\tau_{p_{n(t)}}(t+1) = \infty$ and the message mean to $p_{n(t)}(t+1) = 0$.

\begin{figure}[htb]
\centerline{\includegraphics[width=0.48\textwidth]{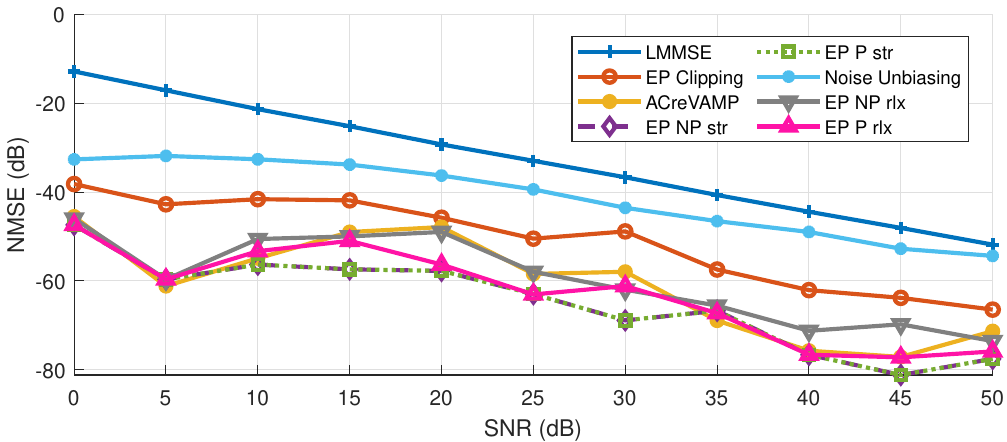}}
\vspace{-3mm}
\caption{Sparse Signal Recovery}
\label{Fig_MSEsolutions}
\vspace{-3mm}
\end{figure}

\begin{figure}[htb]
\centerline{\includegraphics[width=0.48\textwidth]{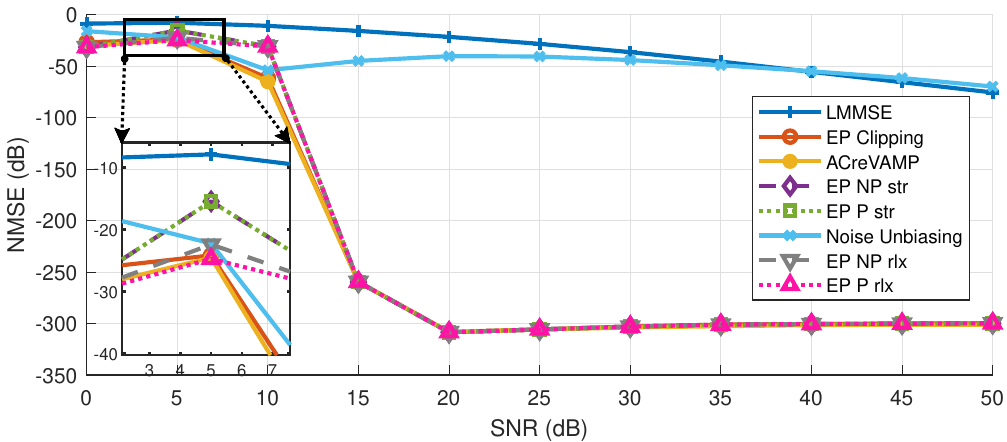}}
\vspace{-3mm}
\caption{BPSK Signal Recovery}
\label{Fig_MSEdddns}
\vspace{-3mm}
\end{figure}

The sparse signal recovery results are shown in Fig.~\ref{Fig_MSEsolutions}, while the BPSK signal recovery results are shown in Fig.~\ref{Fig_MSEdddns}.

In the plots, "EP P" and "EP NP" denote persistent and non-persistent approaches proposed in Section~\ref{sec:asilomarVII}. The labels "str" and "rlx" indicate whether strict checking or relaxed checking, as discussed in Section \ref{sec:asilomarVII}, is used.

From the two plots, we observe that when strict checking is employed, the persistent and non-persistent methods proposed in Section~\ref{sec:asilomarVII} share the same stationary points, as indicated by the overlapping curves. In contrast, under relaxed checking, the persistent and non-persistent approaches converge to different solutions.

As the signal $\bmx$ transitions from continuous-valued to essentially discrete, the proposed ACreVAMP method in Section~\ref{sect:jdfuab} outperforms both the non-persistent and persistent approaches, particularly at low SNR.


\section{Concluding Remarks}\vspace{-1mm}

In this paper, we reviewed the EP algorithm and reVAMP, an application of EP to linear models. EP and reVAMP approximate marginal distributions—referred to as beliefs—through the exchange of messages. While the beliefs must be proper probability distributions normalized to one, the messages themselves may have infinite integral values and therefore need not be proper distributions.

In our setting, due to the use of Gaussian projection, all messages take a Gaussian form. Consequently, when a message has an infinite integral value, it appears as though it has a "negative" variance. Such messages can sometimes prevent the EP/reVAMP algorithm from proceeding. Depending on how these "negative-variance" messages are handled, we proposed three approaches. The non-persistent and persistent EP variants allow the existence of negative message variances, whereas ACreVAMP suppresses or prevents messages with negative variances.

Simulation results demonstrate that the proposed approaches outperform standard benchmark methods. For continuous-valued signal recovery, allowing negative-variance messages leads to superior performance, while for discrete signal recovery, preventing or suppressing negative variances proves more advantageous.

\section{Acknowledgements}\vspace{-1mm}

EURECOM's research is partially supported by its industrial members:
ORANGE, BMW, SAP, iABG,  Norton LifeLock, by the Franco-German projects CellFree6G and 5G-OPERA, the French PEPR-5G projects PERSEUS and YACARI, the EU H2030 project CONVERGE, and by a Huawei France funded Chair towards Future Wireless Networks.

\nocite{zhang2021unifying}
\nocite{heskes2005approximate}
\nocite{wainwright2008graphical}
\nocite{murphy2013loopy}
\nocite{minka2005divergence}
\nocite{zou2022concise}
\nocite{rangan2016fixed}
\nocite{Triki:asilo05}
\nocite{huemer2014cwcu}
\nocite{9928759}

\bibliographystyle{IEEEtran}

\bibliography{asilomar23,SBL_ref2}

\end{document}